\documentclass[11pt]{article}

\usepackage[utf8]{inputenc}
\usepackage[T1]{fontenc}

\usepackage{arxiv}

\usepackage{natbib}
\usepackage{url}
\usepackage{booktabs}
\usepackage{array}
\usepackage{tabularx}
\usepackage{amsmath,amssymb}
\usepackage{graphicx}
\usepackage{xcolor}
\usepackage{tcolorbox}
\tcbuselibrary{skins,breakable}
\usepackage{caption}
\usepackage{hyperref}
\hypersetup{
  colorlinks=true,
  linkcolor=arxivaccent,
  citecolor=arxivaccent,
  urlcolor=arxivaccent,
  filecolor=arxivaccent,
  breaklinks=true,
}

\graphicspath{{figures/}}

\newcommand{\fire}{\texttt{<END\_SPEECH>}}
\newcommand{\eager}{\texttt{<EAGER\_END\_SPEECH>}}
\newcommand{\ctx}{\texttt{<CTX>}}

\title{The Trade-off Was in the Labels: Causal Supervision for Turn-Aware Streaming ASR}

\author{%
  \begin{tabular}{@{}c@{\hspace{4em}}c@{}}
    Bojie Li & Noah Shi \\
    Pine AI & University of Washington
  \end{tabular}%
}
\date{}
\runningtitle{The Trade-off Was in the Labels: Causal Supervision for Turn-Aware Streaming ASR}

\begin{document}
\maketitle

\begin{abstract}
A voice agent must decide, moment to moment, whether the user has finished; silence rarely settles it: a caller reading a phone number pauses mid-digits, a one-word ``Stop!'' ends a turn, a long question carries pauses longer than real turn-gaps. A voice-activity detector plus a silence timeout (the deployed default) cannot separate these, because within-turn pauses routinely exceed between-turn gaps; what distinguishes them is whether the words so far form a complete thought: what a recognizer computes to produce a transcript. We present the first open training recipe and benchmark for \emph{turn-aware} streaming ASR: a small LoRA adapter on Qwen3-ASR-0.6B, trained in hours on one GPU, that transcribes, detects end-of-turn from meaning and silence, handles dictation, and grounds transcription in context. On a deployment-matched benchmark it reaches 0.97 boundary recall at 0.39\,s median latency with 0.3 false fires per speech-minute, replicated on a fresh test set; no silence timeout reaches this point. The recipe rests on one principle: every streaming-decision label must be computable from input up to the decision point. Offline corpora violate it, encoding the future; such \emph{clairvoyant} labels manufactured oscillation and a phantom recall-versus-precision trade-off, exposed when one appended second of silence raised a ``broken'' model's end-of-turn recall from 0.10 to 1.00. The same leak recurred with context: an always-matching biasing prefix became a copied shortcut (40\% intrusion), and counterfactuals disagreeing with the audio cut this to 0.8\% while keeping most of a $+28.9$\,pp entity-recall benefit.
\end{abstract}

\vspace{-0.3em}
\begin{center}
\small
Code: \href{https://github.com/19PINE-AI/turn-aware-asr}{\texttt{github.com/19PINE-AI/turn-aware-asr}} \;\textbullet\; Website: \href{https://01.me/research/turn-aware-asr}{\texttt{01.me/research/turn-aware-asr}}
\end{center}
\vspace{-0.5em}

{\setlength{\intextsep}{6pt plus 2pt minus 2pt}%
\begin{figure}[h!]
  \centering
  \includegraphics[width=\linewidth]{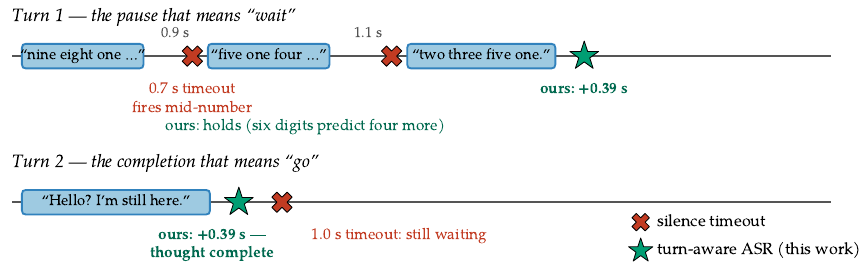}
  \caption{\textbf{No timeout gets both turns right.} At 0.7\,s it interrupts the dictated number twice; at 1.0\,s it still keeps the user waiting after ``I'm still here.'' Ours holds through the pauses and fires 0.39\,s after the thought completes; Figure~\ref{fig:scenarios} gives the full three-scenario setup.}
  \label{fig:teaser}
\end{figure}}

\section{Introduction}
\label{sec:intro}

A voice agent that cannot tell when you have finished speaking fails in two symmetric ways: it interrupts you mid-thought, or it stares at you after you stop. Three ordinary turns show why a single silence threshold cannot avoid both. A caller reading back a phone number pauses between digit groups, ``five five five'' \emph{(pause)} ``oh one two three'', and any timeout short enough to feel responsive fires inside the number, a failure production teams report verbatim \citep{livekit2026eou}. A user who cuts in with ``Stop!'' has ended a turn in a single word, so a timeout patient enough to survive that digit-group pause now leaves an urgent one-word command hanging. A caller working through a long, technical question threads clause-length pauses across half a minute of speech, each one longer than the gap a fast timeout allows, so the agent barges into the question again and again. No one threshold is quick for the one-word turn and patient for the other two, and this is not a matter of tuning: humans exchange turns with median gaps near 200\,ms \citep{stivers2009universals,levinson2015timing}, deployed timeouts sit at 500--1000\,ms \citep{skantze2021review}, and the pauses \emph{within} turns routinely run longer than the gaps \emph{between} them \citep{raux2008optimizing}. What tells a pause from an ending is not its length but whether the words so far form a complete thought, and the words live inside the recognizer.

The conclusion this forces is to move the decision \emph{inside} the recognizer: a single model that, chunk by chunk, transcribes and decides from meaning and silence together whether the turn is over. We call this class \textbf{turn-aware streaming ASR}. The recognizer is the right home for the decision because the completeness signal is a by-product of transcription (a decoder trained to predict the next word already carries whether the utterance can end), so a bolted-on downstream classifier only re-derives, a beat late and from a lossier view, what the recognizer already computes. Frontier voice systems have converged on the same design point, but none discloses how such a model is trained, and the open ecosystem supplies only fragments and no recipe (\S\ref{sec:related}).

\begin{figure}[t]
  \centering
  \includegraphics[width=0.9\linewidth]{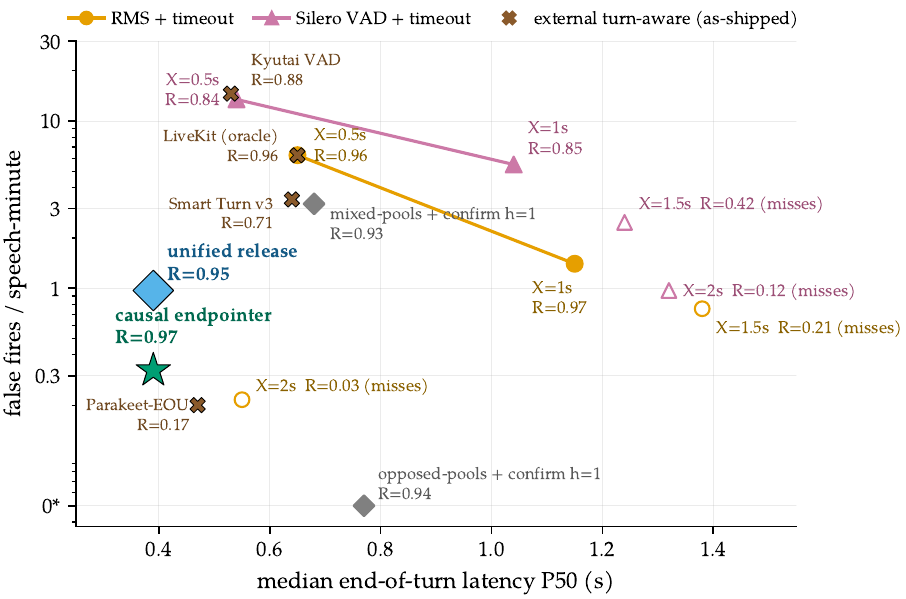}
  \caption{\textbf{The causal model escapes the timeout curve.} Median end-of-turn latency vs.\ false fires (log scale; $0^{*}$ plotted at $0.05$) on identical audio, detector, and scoring. A silence timeout can only slide along its curve; the causally supervised model sits strictly inside the whole family because it reads completeness, firing 0.39\,s after a finished thought while holding through a 1.4\,s mid-sentence pause.}
  \label{fig:tradeoff}
\end{figure}

This paper presents, to our knowledge, the first open, complete training recipe for the class. The system is deliberately small: a LoRA adapter \citep{hu2022lora} on Qwen3-ASR-0.6B \citep{qwen3asr2026}, trained in hours on one GPU from roughly twenty thousand synthesized examples. One checkpoint transcribes incrementally, emits an in-transcript end-of-turn token, holds through dictated digit strings, and grounds transcription in a session context prefix. On a deployment-matched streaming benchmark, and again on a fresh test set drawn after development ended, it reaches an operating point that no silence timeout and none of the open turn-aware systems attain: sub-half-second median latency with higher boundary recall and fewer false fires (Figure~\ref{fig:tradeoff}). The same prefix substantially lifts entity recall on entity-dense audio without degrading transcription quality. The recipe and benchmark are released.

The recipe's substance is not architectural: the entire turn capability comes from the labels, and the paper's central finding is what those labels must satisfy. Supervision inherited from offline corpora defines end-of-turn using audio \emph{after} the decision point, audio a streaming model can never see: offline clips are cut by forced alignment at exactly the boundaries a streaming model must detect, and offline transcripts record who spoke next. We call such labels \emph{clairvoyant}.

Clairvoyant labels manufacture trade-offs that are not real. Our development record holds eight successive \emph{supervision compositions}: the same architecture, adapter, and data volume, relabeled eight ways. Together they traced what looked like a fundamental recall-versus-precision frontier, oscillating between checkpoints that fired eagerly mid-utterance and checkpoints that barely fired at all. A one-variable intervention exposed the frontier as an artifact: appending one second of silence to the evaluation clips took a ``broken'' checkpoint's fire recall from 0.10 to 1.00. One causal labeling rule (fire iff the words so far are semantically complete and at least 0.3\,s of silence has been observed) plus a minimal-pair construction makes training monotone, with nothing else changed, and yields a single checkpoint that dominates all eight predecessors.

The principle is not specific to time. Trained only on examples where the biasing prefix matched the audio, the model learned to \emph{copy} the prefix rather than listen: a wrong-user profile wrote the wrong entity into the transcript 40\% of the time. The leak has the same shape (a training target that depends on a signal the model cannot trust at decision time) and the same repair: counterfactual examples in which context and audio deliberately disagree and the target follows the audio, which drive intrusion below 1\% while largely preserving the biasing benefit. Two instances of one failure class, cured by the same construction, ground a checklist for any streaming decision trained from offline logs.

Our contributions:
\begin{enumerate}\itemsep2pt
  \item \textbf{One small model that transcribes and takes turns, and the first open recipe for it.} A single labeling rule and two pair constructions yield a model that transcribes, emits in-transcript end-of-turn events, holds through dictation, and grounds transcription in context, beating every silence-timeout setting on a deployment-matched benchmark on which the historical model ranking \emph{inverts}. The checkpoint is a research prototype: trained on ${\sim}20$k synthesized examples, it demonstrates what the recipe produces rather than a production system; the reusable artifacts are the recipe, code, and benchmark.
  \item \textbf{A failure class, with a diagnosis by intervention.} Clairvoyant labels (streaming supervision that depends on post-decision input) manufacture training oscillation and phantom capability trade-offs. A one-variable experiment falsifies the frontier that eight supervision compositions had appeared to trace; a synthetic study isolates the mechanism.
  \item \textbf{Evidence that the principle generalizes.} The contextual analog of the temporal leak, context copying under matched-only training, is cured by the analogous counterfactual construction, folding dictation and context grounding into the same checkpoint. Both instances distill into a checklist for streaming supervision pipelines.
\end{enumerate}

\section{Problem formulation and system overview}
\label{sec:task}

\begin{figure}[t]
  \centering
  \includegraphics[width=\linewidth]{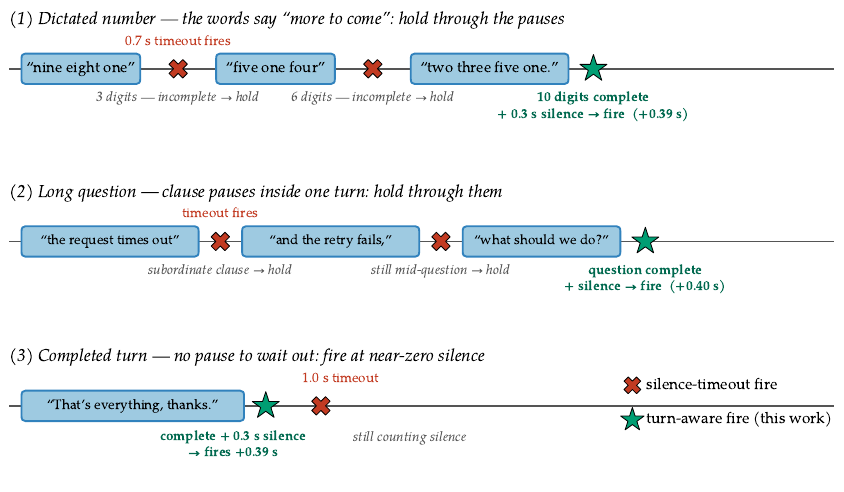}
  \caption{\textbf{The three turns that define turn-aware endpointing, and the read each demands.} Two require \emph{holding through} silence (a dictated number, whose pattern predicts more digits, and a long question, whose clause pauses sit inside a single turn), while a completed utterance requires \emph{firing} at near-zero silence. A fixed silence timeout fires too early on Turns 1--2 and too late on Turn 3; the turn-aware model instead reads, at each 0.5\,s boundary, whether the words so far are complete \emph{and} whether at least 0.3\,s of silence has been observed (the causal rule of \S\ref{sec:recipe}). Schematic timeline; the ``ours'' latencies are the model's measured medians (\S\ref{sec:results}).}
  \label{fig:scenarios}
\end{figure}

\begin{figure}[t]
  \centering
  \includegraphics[width=\linewidth]{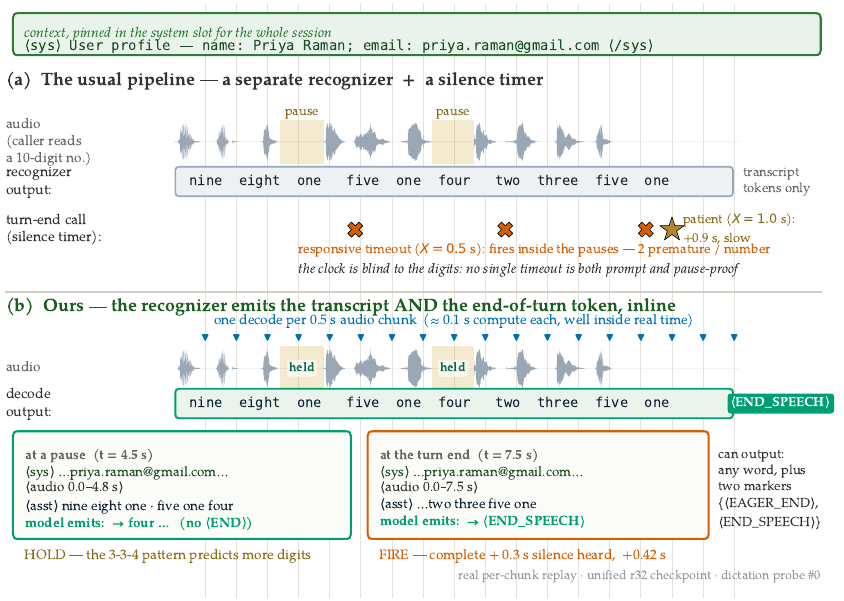}
  \caption{\textbf{How the turn-end decision moves inside the recognizer} (one real per-chunk run on dictation probe~\#0; prompts and outputs verbatim). \textbf{(a)}~The usual pipeline outputs only words and leaves the turn-end call to a downstream silence timer, blind to the digits: a short timeout fires inside the pauses, a long one lags. \textbf{(b)}~Ours emits the end-of-turn marker \emph{in the same decode} as the words, holding through both pauses, then firing 0.42\,s after the last digit. The transcript is identical in both panels; only where the decision is made differs.}
  \label{fig:decodetape}
\end{figure}

Three turns bound the problem, and Figure~\ref{fig:scenarios} shows what each one asks of the model. Two of them require \emph{waiting through} the silence: the dictated number, where six digits of a ten-digit pattern promise more to come, and the long question, where the pauses between clauses sit inside a single unfinished thought. The third, a completed sentence, requires answering \emph{the instant the silence begins}, because a human would reply within roughly 200\,ms \citep{stivers2009universals}. No single silence threshold can serve all three, because what tells them apart is not how long the silence is: on our conversational benchmark the pause lengths inside turns and the gaps between turns overlap almost completely, and the pauses \emph{inside} turns are the longer of the two (\S\ref{sec:analysis}). That overlap, and the dilemma it creates, has defined silence-based turn-taking since \citet{raux2008optimizing} \citep{skantze2021review}. What actually resolves it is whether the words so far form a complete thought: exactly what a recognizer works out on the way to a transcript.

Transcription faces the same kind of problem. A voice agent hears names, stock tickers, and email addresses: words almost impossible to recover from sound alone, yet obvious once you know who is calling and about what. Every major commercial speech-to-text system lets such details be supplied as a hint, yet none publishes controlled numbers on how much they help, and the open model we build on describes its own version in a single qualitative sentence \citep{qwen3asr2026}. A model that accepts such a hint can lean on the caller profile the agent's memory already keeps \citep{li2026uac} to transcribe the whole session more accurately; \S\ref{sec:results} measures exactly that.

\paragraph{What the model reads, and what it decides.} The input is one audio stream, delivered in 0.5\,s slices we call \emph{chunks}. The model must produce two things. One is the running transcript, extended chunk by chunk. The other is the decision this paper is about: the moment the speaker has finished their turn so the agent should reply, classically called \emph{endpointing}. We name the instant the model commits to ``the turn is over'' a \emph{fire}: it cannot be taken back, and every millisecond of delay is felt by the caller. Where this decision is made is the entire design choice. Rather than attach a separate detector to the transcript, we let the recognizer make it. At each chunk the model runs one \emph{decode} (it re-reads the recent audio together with the words it has already emitted and continues the transcript), and the end-of-turn signal is simply one more token it may emit in that same step, drawn from the same vocabulary as the words. A short block of text (for instance, the caller's profile) may be supplied once at the start and stays in view for the whole session. Figure~\ref{fig:decodetape} follows one real decode, chunk by chunk, next to the separate-detector-plus-timeout pipeline it replaces.

\begin{figure}[t]
  \centering
  \includegraphics[width=\linewidth]{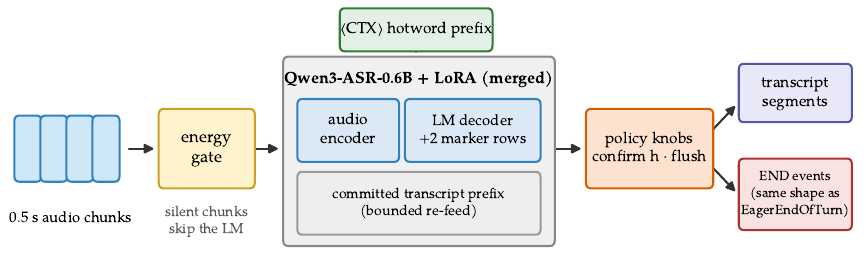}
  \caption{\textbf{System shape.} Audio chunks pass an energy gate; the merged LoRA model transcribes the current bounded segment with the \ctx{} prefix in the system slot and may emit marker tokens; two policy knobs sit behind the model. Outputs are transcript segments plus END events.}
  \label{fig:system}
\end{figure}

\paragraph{What we start from, and what we add.} We build on Qwen3-ASR-0.6B \citep{qwen3asr2026}, an open speech model that transcribes accurately but never signals that the speaker is done. We add that signal as cheaply as possible: we claim two unused entries in the model's token vocabulary as end-of-turn markers, \eager{} and \fire, and lightly fine-tune the model to emit them inside the transcript at the right moments: a LoRA adapter \citep{hu2022lora} of rank 16 for the endpointing experiments and 32 for the full model (details in Appendix~\ref{app:recipe}). Training data is synthesized from LibriSpeech \citep{panayotov2015librispeech} and AMI meeting recordings \citep{carletta2005ami}, and training takes hours on one GPU. Three interpretable dials sit outside the model at run time (Figure~\ref{fig:system}): an \textbf{energy gate} that keeps the model from transcribing during pure silence, a \textbf{confirm horizon} $h$ that waits $h$ further silent chunks before acting on a fire, and a \textbf{max-segment force-flush} that caps how long one transcript segment can grow. Table~\ref{tab:matrix} (Appendix~\ref{app:landscape}) places the system among open and commercial turn-aware systems.

\section{Causal supervision}
\label{sec:method}

\subsection{Clairvoyant labels and the causality principle}
\label{sec:principle}

\begin{figure}[t]
  \centering
  \includegraphics[width=\linewidth]{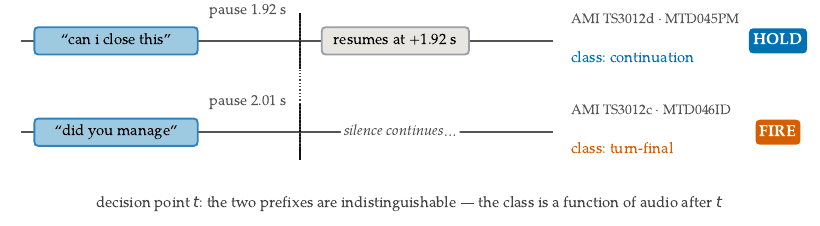}
  \caption{\textbf{Clairvoyant labels.} Two real pauses, 90\,ms apart in length, from held-out AMI stretches of the replay benchmark (texts and gaps verbatim); the label is decided by the future. At the decision point the prefixes are indistinguishable, so the class is a function of audio after $t$, exactly what a clairvoyant label encodes.}
  \label{fig:expauses}
\end{figure}

A streaming model must decide from what it has already heard: at time $t$, only the audio up to $t$ exists. Stated this way the constraint sounds too obvious to violate, yet supervision built from offline corpora violates it systematically, in two places. First, \emph{where the clips end}. Offline ASR corpora are segmented by forced alignment, so every clip stops at the instant the words do, and an endpoint label built on such clips says: fire the moment speech ends. A live channel never presents that condition; when a real speaker stops, silence keeps arriving, and the model must decide during that silence. Second, \emph{who speaks next}. The natural way to label a pause (hold if the same speaker resumes, fire if the turn passes to someone else) reads the answer off the rest of the recording. At the moment of decision, mid-pause, the two cases look identical (\S\ref{sec:analysis}); Figure~\ref{fig:expauses} shows two real pauses, 90\,ms apart in length, whose labels differ only because of what happened afterwards. We give the failure class a name:

\begin{tcolorbox}[breakable,title=\textbf{Definition: clairvoyant labels},colback=red!3!white,colframe=red!50!black]
A label for a streaming decision at time $t$ is \emph{clairvoyant} if its value depends on input after $t$. Clairvoyant labels do not merely add noise: because the dependence is systematic, they split the training data into pools that are each self-consistent yet demand opposite behavior on the same observable input. The symptoms are (a)~training that oscillates between two modes instead of converging, and (b)~apparent capability trade-offs across checkpoints: a Pareto frontier that is not real. \textbf{Principle: every training label and every evaluation target for a streaming decision must be computable from the input up to the decision point.}
\end{tcolorbox}

Clairvoyance is target leakage transposed into time, but the class is wider than time: the same failure appears whenever the target correlates with any signal the model cannot \emph{trust} at decision time. A context prefix that always matches the audio during training is such a signal, because at deployment the profile can simply be wrong (\S\ref{sec:ctxrecipe}; Figure~\ref{fig:twoaxes}, Appendix~\ref{app:labels}, shows the two instances side by side). In both cases the cure is a data construction, not a loss term: build pairs of examples in which the untrustworthy signal is uninformative by design, so the only policy consistent with the training data is the one that keys on observables. \S\ref{sec:analysis} presents the evidence that this, and nothing architectural, was the obstacle: most concretely in a one-variable check (\S\ref{sec:intervention}) where appending a single second of silence to the evaluation lifts a ``broken'' checkpoint from 0.10 to 1.00 fire recall, dissolving the very precision axis the frontier was thought to trade against.

\subsection{Supervising end-of-turn detection}
\label{sec:recipe}

\begin{figure}[t]
  \centering
  \includegraphics[width=\linewidth]{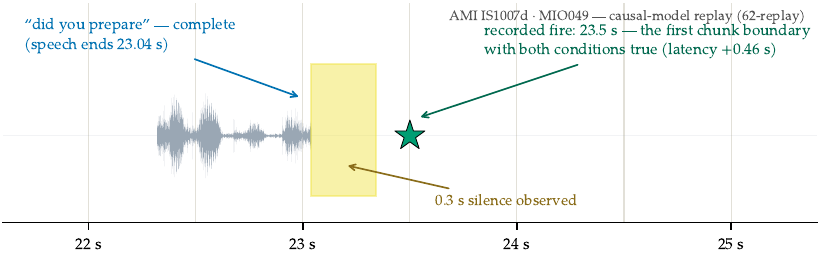}
  \caption{\textbf{Supervising end-of-turn detection.} In deployment-matched replay, ``did you prepare'' is complete at 23.04\,s, 0.3\,s of silence is observed by 23.34\,s, and the recorded fire lands at the next 0.5\,s chunk boundary (latency $+0.46$\,s). Waveform and fire timestamp verbatim from the recorded causal-model replay.}
  \label{fig:exrule}
\end{figure}

The entire turn capability comes from the labels. One rule generates every training example: \emph{emit \eager\fire{} iff the words so far are semantically complete AND at least 0.3\,s of silence has been heard since the last speech.} Both conditions are checkable from the audio prefix alone: completeness by a transcript heuristic (an utterance that trails off in a filler or connective, cuts mid-word, or is too short to be an acknowledgment counts as incomplete; Appendix~\ref{app:labels}), silence from the synthesized waveform itself. Figure~\ref{fig:exrule} shows the rule firing in deployment-matched replay.

Eight example templates, the \emph{schemas} of Table~\ref{tab:schemas}, instantiate the rule. Two constructions among them teach the model to separate meaning from silence. The \textbf{minimal pair} (schemas 1--2; Figure~\ref{fig:minimalpair}, Appendix~\ref{app:labels}) presents the same complete utterance twice: with a silence tail, labeled fire, and without one, labeled hold. The only difference between the two examples is the silence, so completeness alone cannot trigger a fire; the model has to \emph{see} the silence (the direct fix for row 4 of Table~\ref{tab:composition}, where silence had become optional). The \textbf{pause pair} (schemas 4--5) presents two utterances separated by a pause, and whether the pause carries a marker depends only on whether the words before it are complete: ``\ldots and then, \emph{(pause)}'' holds; ``\ldots that's everything. \emph{(pause)}'' fires, even if the speaker later resumes.

The remaining schemas close deployment gaps. A quick resume after a legitimate fire simply becomes two transcript segments, a measured cost rather than an error (\S\ref{sec:bench}): at decision time the correct answer does not exist on the channel. Schemas 7--8 add the silence-heavy audio that utterance corpora never contain, and pair examples mix same- and different-speaker sources, so the decision cannot key on speaker identity, which a single-channel deployment never observes. A size-matched ablation confirms what each construction buys (Appendix~\ref{app:tables}, Table~\ref{tab:ablation}).

\begin{table}[t]
  \centering\small
  \caption{The eight training schemas (${\sim}12$k examples; LibriSpeech and AMI roughly 50/50). $M$ = \eager\fire.}
  \label{tab:schemas}
  \begin{tabular}{lll}
    \toprule
    \# & Construction & Target \\
    \midrule
    1 & complete utterance + 0.3--1.2\,s silence tail & \texttt{text} $M$ \\
    2 & \emph{same utterances as \#1}, tail ${\leq}0.1$\,s & \texttt{text} \\
    3 & speech truncated mid-utterance at 30--80\% & partial \texttt{text} \\
    4 & complete A + gap 0.3--2.5\,s + B (+ tail) & \texttt{A} $M$ \texttt{B} $M$ / \texttt{A} $M$ \texttt{B} \\
    5 & \emph{incomplete} A + gap + B + tail & \texttt{A B} $M$ \\
    6 & complete utterance + long silence (2--4\,s) & \texttt{text} $M$ \\
    7 & pure silence 0.5--4\,s & (empty) \\
    8 & leading silence 0.5--2\,s + utterance + tail & \texttt{text} $M$ \\
    \bottomrule
  \end{tabular}
\end{table}

\subsection{Supervising dictation and context biasing}
\label{sec:ctxrecipe}

\begin{figure}[t]
  \centering
  \includegraphics[width=\linewidth]{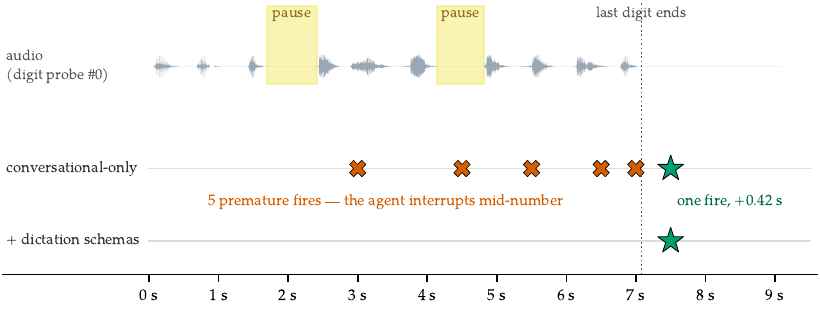}
  \caption{\textbf{Supervising dictation.} Probe item 0, identical audio under two supervisions: the conversational-only model reads each digit group as a finished thought and fires five times inside the number; the dictation schemas hold through both pauses and fire once, $+0.42$\,s after the last digit. Fires verbatim from the recorded probe results.}
  \label{fig:exdictation}
\end{figure}

The same discipline extends to the two transcription behaviors of \S\ref{sec:task}. \textbf{Dictation:} a partial phone number followed by a pause is labeled \emph{hold}, because the pattern predicts more digits; a complete number plus observed silence fires. Figure~\ref{fig:exdictation} shows the effect on identical probe audio: five mid-number interruptions become one fire, $+0.42$\,s after the last digit. \textbf{Spelled entities and context:} synthesized utterances spell out uncommon names and email addresses, targeted in normalized written form, half of them with a user profile in the context slot. These schemas extend the pool to ${\sim}15$k examples for retraining the same LoRA.

\begin{figure}[t]
  \centering
  \includegraphics[width=\linewidth]{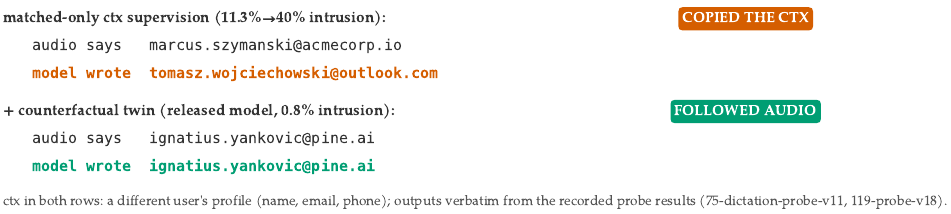}
  \caption{\textbf{Supervising context biasing.} Under a wrong-user profile, the matched-only model writes the profile's email instead of the audio's (copying); the counterfactually trained unified model follows the audio. Outputs from the recorded spelled-entity probes.}
  \label{fig:exctx}
\end{figure}

Training only on matching profiles, however, rebuilds the \S\ref{sec:principle} failure on the context axis. If the profile always agrees with the audio, the model can score perfectly without listening: it learns to copy whatever entity the prefix names, and at deployment a wrong profile writes the wrong entity into the transcript (Figure~\ref{fig:exctx}; \S\ref{sec:results} measures the damage).

The \textbf{counterfactual twin} removes the shortcut. A third of the spelled examples carry a \emph{conflicting} profile: the context names one identity, the audio spells another, and the target follows the audio, the same construction as the minimal pair. A second counterfactual of the same shape covers natural-speech biasing (an entity list that disagrees with the audio; the target again follows the audio). The unified pool (${\sim}20$k examples) folds in both counterfactuals plus a 20\% plain-transcription replay fraction that anchors offline WER (Appendix~\ref{app:recipe}); the unified adapter is rank 32, the capacity \S\ref{sec:results} shows the added breadth requires.

\section{Evaluation protocol}
\label{sec:eval}

\subsection{Streaming replay benchmark}
\label{sec:bench}

Endpointing is evaluated by \emph{streaming replay}: recorded audio is streamed through the live system chunk by chunk, exactly as a deployment would deliver it. The causality principle applies to the benchmark itself. An evaluation whose clips end at forced-alignment cuts demands behavior on a condition deployment never produces, and \S\ref{sec:analysis} shows that such an evaluation actively misdirects development.

\paragraph{Material and ground truth.} Continuous single-channel stretches from held-out AMI meetings: one target speaker's utterances at their true timeline offsets, silence between them, 30--60\,s per stretch. This reproduces the one condition that drives endpointing in a live voice agent: the audio is continuous and never ends at a speech boundary. Each utterance-end boundary is classified by what is observable \emph{on this channel}: \textbf{turn-final} (same-speaker gap ${\geq}2$\,s), where the system should fire; \textbf{continuation} (gap in $[0.3,2)$\,s, same speaker resumes), where firing is a measured segmentation cost rather than an error, because the correct answer depends on the future; and \textbf{internal} (pauses inside utterances, or ${<}0.3$\,s after speech), where firing is a false fire. Metrics: boundary recall within $[-0.25,+1.5]$\,s of the true end, false fires per speech-minute, latency percentiles, the resume-after-fire rate (how often a fire lands at a continuation boundary), streaming WER of the flushed transcripts, and per-chunk compute. The development set has 25 stretches; a twice-larger confirmation set was drawn after all development ended, and a 100-stretch held-out set is used for the unified model. An early version of the benchmark itself contained a clairvoyant rule, caught by our own checklist; all numbers use the corrected classification (Appendix~\ref{app:eval}).

\paragraph{Why comparisons run behind the energy gate.} Models trained only on speech-initial examples hallucinate markers on silence at roughly two fires per second, so recall without the gate is uninterpretable: a model firing randomly at 1.9\,Hz scores 0.96 against the tolerance window (Appendix~\ref{app:eval}, Figure~\ref{fig:silence}). All comparisons therefore run behind the energy gate; the causal recipe additionally bakes silence schemas into training.

\subsection{Probes and offline evaluations}
\label{sec:probes}

Each \S\ref{sec:task} scenario has a dedicated probe, and both run through the exact streaming stack of \S\ref{sec:bench}. The \textbf{dictation probe} replays ten-digit phone numbers (3-3-4 groups from held-out Free Spoken Digit Dataset speakers \citep{jackson2018fsdd}, with inter-group pauses) and scores premature fires per number, final-boundary recall, and digit accuracy. The \textbf{spelled-entity probe} decodes synthesized utterances spelling uncommon names and email addresses, with and without a user-profile prefix, plus a wrong-user \emph{distractor} profile, and scores exact match and wrong-profile intrusion. Both probes were enlarged after development ended; unified-model numbers use the enlarged versions. \textbf{Context biasing} on natural speech is measured on Earnings-22 \citep{delrio2022earnings22}, spontaneous earnings-call speech dense in tickers, executives, and products, with entities LLM-extracted and filtered to genuine proper nouns, as entity recall with and without a relevant hotword prefix. \textbf{Offline WER} is scored on the full official LibriSpeech splits, and \textbf{serving} is measured on stock vLLM rather than a simulator (Appendix~\ref{app:serving}).

\section{Results}
\label{sec:results}

\subsection{End-of-turn detection}
\label{sec:endpointresults}

\begin{table}[t]
  \centering\small
  \caption{\textbf{Main endpointing result} (deployment-matched replay, energy gate on; the latency-vs-false-fire view is Figure~\ref{fig:tradeoff}). Rows 1--4: the 25-stretch development benchmark. Last row: the fresh twice-larger confirmation set drawn after all development ended. WER is the per-stretch \emph{median} streaming WER; the corresponding means, dominated by long-monologue repetition loops when no force-flush bounds the segment, are in Appendix~\ref{app:tables}, and the force-flush of Appendix~\ref{app:serving} bounds that tail.}
  \label{tab:main}
  \begin{tabular}{lccccc}
    \toprule
    Policy & Recall\,$\uparrow$ & P50\,$\downarrow$ & P95\,$\downarrow$ & False/min\,$\downarrow$ & WER$_\text{med}$\,$\downarrow$ \\
    \midrule
    mixed-pools + gate + confirm $h{=}1$ & 0.927 & 0.68\,s & 1.27\,s & 3.2 & 0.31 \\
    opposed-pools + gate + confirm $h{=}1$ & 0.938 & 0.77\,s & 1.06\,s & 0.0 & 0.27 \\
    \textbf{causal + gate (ours)} & \textbf{0.969} & \textbf{0.39\,s} & \textbf{0.70\,s} & \textbf{0.3} & 0.28 \\
    causal + gate + confirm $h{=}1$ & 0.969 & 0.89\,s & 1.20\,s & 0.0 & 0.28 \\
    causal + gate, fresh 50-stretch set & 0.924 & 0.42\,s & 0.70\,s & 0.2 & 0.30 \\
    \bottomrule
  \end{tabular}
\end{table}

Table~\ref{tab:main} reports the main result. The baselines are the two strongest earlier supervision compositions from the development record of \S\ref{sec:analysis}, \emph{mixed-pools} (row 3 of Table~\ref{tab:composition}) and \emph{opposed-pools} (row 6), each at its best knob setting. The causal model with the gate alone is better on every axis at once: its \emph{raw} false-fire rate beats the mixed-pools model's \emph{confirmed} rate, at half the latency and higher recall. Against the deployed cascade the comparison is structural. A timeout must be long enough to survive within-turn pauses, so it can only trade latency against false fires along one curve; the causal model escapes the curve by reading completeness, combining exactly the pair of behaviors \S\ref{sec:task} argued no threshold can combine. Sweeping the full timeout family over two detectors and all settings on both stretch sets, no setting reaches its latency-and-false-fire corner at any recall (Appendix~\ref{app:tables}). The model also no longer relies on inference-time suppression: the confirm horizon cancels only 3 fire candidates for the causal model versus 241 for the mixed-pools model, so the phrase-versus-turn discrimination moved from inference policy into the weights. The result survives fresh data (Table~\ref{tab:main}, last row): latency and false fires hold, and the small recall dip is consistent with sampling variation (Appendix~\ref{app:tables}). By design, a quick resume after a legitimate fire yields two transcript segments; deployments that prefer merged segments pay $+0.5$\,s via the confirm horizon.

\paragraph{External turn-aware systems on the same protocol.} We ran the open turn-aware systems of \S\ref{sec:related} through the identical benchmark as-shipped, sweeping their public thresholds rather than retraining (Figure~\ref{fig:tradeoff}). None reaches the causal model's corner: the standalone detectors are gated on VAD silence by construction and slide along the timeout curve, LiveKit's text detector reaches high recall only with an oracle transcript and many times our false-fire rate, and the open recognizers with turn signals either rarely fire (Parakeet-EOU) or fire spuriously (Kyutai). To our knowledge this is the first measurement of the open turn-aware class on a common protocol; it is as-shipped, not best-achievable, since none of these systems was tuned for AMI's close-talking microphone audio.

\subsection{Context biasing and dictation}
\label{sec:ctxresults}

\begin{figure}[t]
  \centering
  \includegraphics[width=\linewidth]{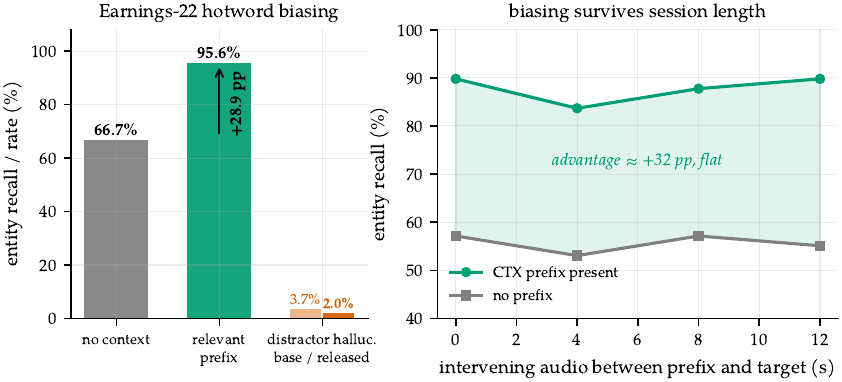}
  \caption{\textbf{Context biasing, measured.} (a)~Earnings-22 base-model entity recall without context and with the relevant hotword prefix ($+28.9$\,pp), and the distractor-prefix hallucination rate for the base model versus the unified rank-32 model, whose natural-speech counterfactual drives it from 3.7\% to 2.0\%. (b)~Recall vs.\ seconds of unrelated speech between prefix and target: the advantage does not decay with session age.}
  \label{fig:biasing}
\end{figure}

\paragraph{Biasing works, and survives session length.} On Earnings-22, a relevant hotword prefix lifts the base model's entity recall from 66.7\% to 95.6\% while nudging WER slightly down: the prefix acts as a domain hint, not a distortion (Figure~\ref{fig:biasing}a). With up to 12\,s of unrelated speech interposed between prefix and target words, the advantage stays flat (Figure~\ref{fig:biasing}b), which is the property a voice agent actually needs, since the user's account entities must help at minute five, not just in the first utterance. The base model's report describes this capability qualitatively without numbers \citep{qwen3asr2026}; to our knowledge these are the first on this model family \citep[cf.][]{contextualearnings2026}.

\paragraph{Dictation, and the copying result.} The conversational-only endpointing model of \S\ref{sec:endpointresults} fails both \S\ref{sec:probes} probes exactly as \S\ref{sec:task} predicts: it reads a digit group as a finished thought, firing prematurely five times per dictated number, and spells emails at 4\% exact match (Table~\ref{tab:dictation}). The dictation schemas of \S\ref{sec:ctxrecipe} fix the former immediately. The context schemas, trained with matched profiles only, produce the predicted copying: the model writes whatever entity the prefix names, so a wrong-user profile puts the wrong email into the transcript, and the intrusion rate grows from 11.3\% to 40\% as training lengthens. Adding the counterfactual twin drives intrusion to zero at every checkpoint on the training probe, and to 0.8\% (95\% CI $[0.2,3.0]$\%) on the enlarged post-development probe, while leaving dictation and matching-context accuracy intact (Table~\ref{tab:dictation}). Context becomes a prior, not a substitute for listening.

\begin{table}[t]
  \centering\small
  \caption{\textbf{Dictation and spelled-entity probes.} Premature fires per dictated 10-digit number, final-boundary recall, digit accuracy; exact-match on spelled names/emails without/with the profile; wrong-profile intrusion. Conversational-only and unified rows are scored on enlarged probes; the intermediate row uses the smaller original probes. Confidence intervals and a scoring-window sensitivity analysis are in Appendix~\ref{app:tables}. The conversational-only model's nonzero intrusion is the base model's native context-following, which that fine-tune leaves untouched.}
  \label{tab:dictation}
  \setlength{\tabcolsep}{2.6pt}
  \footnotesize
  \begin{tabular}{lccc|ccc}
    \toprule
    & Premat./seq\,$\downarrow$ & Final rec.\,$\uparrow$ & Digit acc\,$\uparrow$ & Name ${-}/{+}$ & Email ${-}/{+}$ & Intrus.\,$\downarrow$ \\
    \midrule
    conversational only & 4.96 & 0.82 & 0.889 & 0.25\,/\,0.82 & 0.04\,/\,0.58 & 1.7\% \\
    timeout $X{=}0.5$\,s & 2.00 & 1.00 & -- & -- & -- & -- \\
    timeout $X{=}1.0$\,s & 0.74 & 1.00 & -- & -- & -- & -- \\
    $+$ dictation (match-ctx) & 0.36 & 0.78 & 0.996 & 1.00\,/\,1.00 & 0.03\,/\,0.93 & 11.3\% \\
    \textbf{unified (r32)} & \textbf{0.16} & \textbf{0.88} & \textbf{0.998} & \textbf{0.98\,/\,1.00} & 0.11\,/\,\textbf{0.93} & \textbf{0.8\%} \\
    \bottomrule
  \end{tabular}
\end{table}

\subsection{The unified model and its costs}
\label{sec:released}

The release is the label-spec builder and training code, the probes, and the replay benchmark. The unified checkpoint, a rank-32 adapter trained on one pool of ${\sim}20$k examples, detects end-of-turn in conversation, holds through dictation better than every timeout setting, grounds spelled entities in context, and ignores a wrong profile (Table~\ref{tab:dictation}, last row), with no deployment-time checkpoint swap; on the 100-stretch held-out replay set it scores 0.982 boundary recall at 1.30 false fires per speech-minute, still strictly inside the timeout family of Figure~\ref{fig:tradeoff}. Its endpointing numbers across stretch sets, the confirm-horizon dial, a scoring-window sensitivity analysis, and three-seed retraining stability are collected in Appendix~\ref{app:tables}. We stress that this checkpoint is a research prototype: trained on ${\sim}20$k synthesized examples, it is meant to demonstrate the recipe, not to serve production traffic, and the enduring contribution is the recipe and findings rather than the weights themselves.

\paragraph{Which trade-offs are real.} Folding four behaviors into one adapter costs endpointing precision: 0.97 false fires per speech-minute on the development set versus 0.3 for the pure-endpointing checkpoint. This is a genuine capacity trade-off, which we report and do not tune away; a deployment that only needs turn-taking can run the pure checkpoint. It is also the one trade-off that survives the diagnosis. The recall-versus-precision frontier dissolved when the labels became causal (\S\ref{sec:analysis}); the breadth-versus-precision cost at fixed adapter capacity did not, but it recedes as capacity grows, which is why the unified model uses rank 32: at that rank the added replay and biasing schemas leave endpointing recall essentially intact, as they did not at rank 16. The context axis carries the same lesson. The spelled-entity counterfactual alone still leaves natural-speech distractors hallucinated 6.3\% of the time; folding the natural-speech counterfactual into the unified pool cuts this to 2.0\%, below the base model's 3.7\%, while preserving nearly all of the biasing uplift.

\paragraph{Costs.} The endpoint fine-tune costs $+1.1$/$+2.7$\,pp offline WER on LibriSpeech clean/other. Banning the marker tokens at decode time shows the markers themselves are not the cause; the regression comes from fine-tuning on a narrow synthetic distribution, and the plain-transcription replay fraction in the unified model recovers most of it (Appendix~\ref{app:recipe}, Table~\ref{tab:wer}). Streaming WER, the deployment-relevant number, is unaffected. Serving runs on stock vLLM \citep{kwon2023vllm} within real-time budgets through 16 concurrent sessions on a fractional-GPU slice. The serving primitive is simple: re-feed the last bounded window of audio each chunk, with a force-flush bounding the WER and compute tails (details and caveats in Appendix~\ref{app:serving}). An in-engine efficiency variant, windowed KV with the \ctx{} prefix pinned as an attention sink \citep{xiao2024streamingllm}, is validated in a companion systems paper \citep{li2026metronome}.

\section{Analysis}
\label{sec:analysis}

The claim that the labels are the load-bearing element of \S\ref{sec:method} rests on a controlled record: eight supervision compositions sharing architecture, adapter, data volume, and source corpora, differing only in the labels. This section presents the record, the intervention that diagnosed it, the mechanism behind the oscillation, and a synthetic study that isolates the mechanism outside speech.

\subsection{Supervision-composition study}
\label{sec:record}

\begin{figure}[t]
  \centering
  \includegraphics[width=\linewidth]{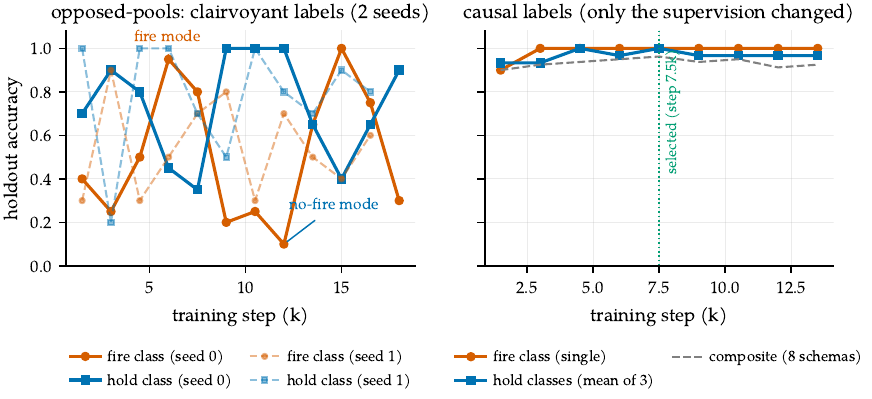}
  \caption{\textbf{The fingerprint of contradictory supervision.} Left: under the opposed-pools composition, holdout accuracy oscillates for the entire run between a fire mode and a no-fire mode, reproducibly across two training seeds (solid, dashed); the oscillation is a property of the labels, not of a seed. Right: the causal recipe of \S\ref{sec:recipe}, with the same architecture, adapter, and data volume and only the labels changed, climbs monotonically, with both classes at ceiling simultaneously from step 3k on.}
  \label{fig:oscillation}
\end{figure}

Table~\ref{tab:composition} compresses the record. The first composition was supervised on offline clips alone (fire at every clip end); it scored perfectly offline and was unusable live, firing 89 times per speech-minute mid-utterance. Successive compositions added pools intended to fix this, and every addition moved the model \emph{along} an apparent recall-versus-precision frontier, never off it. Two behaviors recurred across the family. Checkpoints were bimodal: eager ones fired mid-utterance, conservative ones missed turns, and no checkpoint did both jobs. And late compositions did not converge at all: holdout accuracy oscillated for the entire run between a \emph{fire mode} and a \emph{no-fire mode}, as in Figure~\ref{fig:oscillation} (left). By the eighth composition (row 6) the working conclusion, recorded in the project log at the time, was that the frontier was fundamental, with a recommendation to ship two checkpoints, one eager and one conservative.

\begin{table}[t]
  \centering\small
  \caption{\textbf{The supervision-composition record.} Eight compositions, one architecture, compressed into six rows: rows 1 and 2 each span two iterations (a data scale-up; a truncation-dose increase), and each later row adds one labeled pool to the mix above it, except row 6, which \emph{replaces} row 5's same-audio hold pool with disjoint audio. ``Early'' latencies are negative: the model fires seconds before the turn ends. Row 7 is the causal recipe of \S\ref{sec:recipe}.}
  \label{tab:composition}
  \setlength{\tabcolsep}{4pt}
  \begin{tabularx}{\textwidth}{@{}c>{\raggedright\arraybackslash}X>{\raggedright\arraybackslash}X@{}}
    \toprule
    \# & Supervision change (cumulative) & Observed behavior \\
    \midrule
    1 & offline endpoint pool only: fire at every clip end & fires eagerly and ``accurately'' offline (100\% fire recall); in streaming replay, fires 89$\times$ per speech-minute mid-utterance \\
    2 & $+$ truncated-speech hold pool (11\%${\to}$38\% of mix) & diminishing returns: median fire timing improves only $-13.1{\to}-9.6{\to}-7.6$\,s (still seconds early) as the dose rises \\
    3 & $+$ trailing-silence fire pool & streaming latency crosses zero (P50 $+0.32$\,s) but offline fire recall regresses $100{\to}82$\%; the ``trade-off'' is first declared \\
    4 & $+$ complete utterances with \emph{and} without silence tails, both labeled fire & silence becomes optional again; a strictly intermediate point on the same frontier; frontier declared fundamental \\
    5 & $+$ hold pool on the \emph{same} audio as existing fire examples & collapse: opposite labels on identical inputs; the model stops firing on meeting audio entirely (fire recall 0\%) \\
    6 & row 5's hold pool rebuilt on \emph{disjoint but identically distributed} audio & oscillation between a fire mode and a no-fire mode for the whole run (Figure~\ref{fig:oscillation}, left); no checkpoint dominates; two-checkpoint ship recommended \\
    \midrule
    7 & \textbf{causal relabel (\S\ref{sec:recipe}):} fire iff complete \emph{and} ${\ge}0.3$\,s observed silence & monotone training, both classes at ceiling simultaneously; single checkpoint dominates every predecessor (\S\ref{sec:results}) \\
    \bottomrule
  \end{tabularx}
\end{table}

\subsection{Diagnosis by intervention}
\label{sec:intervention}

\paragraph{The experiment.} One change to the \emph{evaluation}, and none to any model, collapsed the frontier's fire-recall axis: we appended 1.0\,s of silence to each clip of the legacy offline benchmark and re-scored the historical checkpoints (Table~\ref{tab:silappend}). The ``broken'' opposed-pools model's fire recall rose from 0.10 to 1.00 and the mixed-pools model's from 0.82 to 0.98, while the offline-labeled composition, which never depended on hearing silence, was the unchanged control. Under deployment-realistic input, all three fire essentially perfectly on complete utterances. The capability axis that four compositions of data engineering had been fighting was an artifact of where the evaluation clips ended.

\begin{table}[h]
  \centering\small
  \caption{\textbf{The silence-append intervention.} Fire recall on complete single utterances from the legacy offline benchmark, before and after appending 1.0\,s of silence to each clip.}
  \label{tab:silappend}
  \begin{tabular}{lccc}
    \toprule
    Model (supervision composition) & original clips & $+1.0$\,s silence & $\Delta$ \\
    \midrule
    offline-labeled (rows 1--2) & 1.00 & 1.00 & 0.00 \\
    mixed-pools (row 3) & 0.82 & 0.98 & $+0.16$ \\
    opposed-pools (row 6) & 0.10 & 1.00 & $+0.90$ \\
    \bottomrule
  \end{tabular}
\end{table}

\paragraph{The diagnosis.} Why did the clips end where they did? Offline ASR datasets are segmented by forced alignment, so clips are cut at the end of speech, and evaluation clips inherit the cut. An offline endpoint benchmark built on such clips therefore demands a fire on audio that ends \emph{during or immediately after} speech, a condition deployment never presents. The streaming pools taught the opposite, correctly: never fire before silence is observed. The condition ``complete utterance, no trailing silence yet'' thus carried the label \emph{fire} in one pool and \emph{hold} in the other; in the eighth composition this was 50/50 label noise on the exact decision the model exists to make. A second mechanism had the same structure: the historical pools separated ``disfluent pause, hold'' from ``turn boundary, fire'' using the transcript's segmentation. Measured at the decision point, mid-pause, the two classes are statistically indistinguishable: their silence-gap distributions overlap almost completely, and the disfluent pauses are the longer ones. The label is a function of who spoke \emph{next}, that is, of the future. Both mechanisms are exactly the clairvoyance of \S\ref{sec:principle}, and nothing about either is specific to speech; Figure~\ref{fig:contradictions} (Appendix~\ref{app:markerprob}) illustrates both leaks.

\paragraph{Why oscillation, specifically.} Rows 5 and 6 of Table~\ref{tab:composition} form a natural two-condition experiment, differing only in where the contradictory hold labels sit. When opposite labels sit on literally identical audio (row 5), the conflict is visible to the loss on every example, and the model settles deterministically on one side: it simply stops firing. When opposite labels sit on disjoint but identically distributed audio (row 6), no single example exposes the conflict; each pool is separately learnable, incompatible only in aggregate, and training cycles between two modes that each satisfy one pool. The oscillation is not instability to be scheduled away; it is the loss surface reporting a contradiction that no individual example states. A probe of the model's output probabilities (Appendix~\ref{app:markerprob}, Figure~\ref{fig:markerprob}) locates the cycling in the weights rather than at a decision threshold: the probability of committing to the marker swings between snapshots in lockstep with the accuracy, and no held-out example hovers near the decision boundary.

\paragraph{The evaluation was steering development.} The legacy offline benchmark did not just mis-score the historical models; it steered four compositions of data engineering toward a frontier that does not exist, and the deployment-matched protocol of \S\ref{sec:bench} re-\emph{ranks} them (Appendix~\ref{app:eval}, Figure~\ref{fig:inversion}): the offline champion is the worst live model, its offline strength being a premature-fire pathology, while the opposed-pools model, scored at 10\% offline recall, is the best prior endpointer. Teams selecting turn-aware checkpoints on offline evaluations whose clips end at the speech boundary should expect the same inversion.

\subsection{Synthetic isolation of the mechanism}
\label{sec:toy}

The failure reproduces outside speech, with no acoustics and no pretraining (setup in Appendix~\ref{app:toy}). A two-layer, ${\sim}30$k-parameter transformer is trained on a synthetic stream in which completeness is readable from the prefix, each gap's outcome is not, and a tunable fraction $f$ of the labels is clairvoyant by construction. At $f{=}0$ training is textbook monotone. As $f$ grows, the fingerprint develops in order (Figure~\ref{fig:toy}, Appendix~\ref{app:toy}): fire- and hold-class accuracy become anti-correlated; checkpoints spread into an anti-diagonal cloud in the recall-precision plane, an apparent Pareto frontier that is in fact one model swinging, sampled at different steps; and outright mode-flipping begins only at high $f$. Why high $f$ is required is itself informative: the causal label remains the per-prefix \emph{majority} for every $f{<}1$, so a stable single-mode optimum survives until the contradictory pool is large enough to destabilize it. The speech model of \S\ref{sec:record} lived in the high-$f$ regime by construction, because near-balanced opposed pools are exactly what mixing offline supervision with streaming reality produces.

\section{Discussion}
\label{sec:discussion}

\paragraph{A checklist.} The failure mode generalizes to any model that must act mid-stream on offline-labeled data. Four questions for a streaming supervision pipeline:

\begin{tcolorbox}[breakable,colback=blue!3!white,colframe=blue!45!black,title=\textbf{Is your supervision clairvoyant?}]
\begin{enumerate}\itemsep2pt
  \item \textbf{Prefix test.} For each label at decision time $t$: is it computable from input up to $t$? \emph{(Our pause labels were keyed on the next speaker.)}
  \item \textbf{Twin test.} Can two causally identical prefixes receive different labels anywhere in the data? \emph{(Same complete-utterance prefix: fire in one pool, hold in the other.)} Minimal pairs differing in an \emph{observable} are the fix, not the bug.
  \item \textbf{Phantom-condition test.} Does the eval demand behavior on conditions deployment never produces? \emph{(Clips cut at end-of-speech; deployment always delivers the next silence.)}
  \item \textbf{Oscillation test.} Does training bounce between modes that each satisfy part of the data? That is the loss surface reporting a contradiction.
\end{enumerate}
\end{tcolorbox}

\paragraph{Limitations.} Single-speaker, single-channel English; our conversational endpointing evidence comes entirely from held-out AMI meetings, so cross-corpus generalization and multi-speaker mixed-channel audio are future work. The dictation capability transfers zero-shot to shorter enumerations but not to longer-than-trained ones (on 16-digit card numbers the model fires at its learned ten-digit completeness point, transcription unaffected), so a learned completeness judge remains the general mechanism beyond pattern-specific schemas. A symmetric edge sits at the short end that motivated the problem: the completeness heuristic marks sub-three-word utterances incomplete so the model will not fire on backchannels, which means a genuine one-word command such as ``Stop!'' is not guaranteed the near-instant fire a barge-in wants: telling imperatives from acknowledgments is the same learned-completeness problem, sharpened, not a new one. The unified adapter's breadth-versus-precision and recall-versus-hallucination costs are quantified in \S\ref{sec:released}, and a turn-taking-only deployment can run the pure endpointing checkpoint; concurrency numbers are lower bounds measured on a shared GPU under contention (Appendix~\ref{app:serving}). The checkpoint is trained on only ${\sim}20$k synthesized examples on a single GPU: it is a research prototype that validates the recipe, and we make no claim about its production readiness; scaling the same recipe to production-grade data is left to future work.

\section{Related work}
\label{sec:related}

\paragraph{Endpointing and end-of-turn detection.} Classical endpointing thresholds a VAD \citep{silero2021vad} with a silence timeout \citep{raux2008optimizing}; learned refinements predict end-of-query from acoustics \citep{shannon2017improved}, jointly train an endpointer with the recognizer \citep{chang2019joint,li2020towards,bijwadia2023unified}, or distill tiny acoustic end-of-turn models \citep{helwani2026hierarchical,smartturn2025}. Recent systems fuse acoustic and linguistic cues \citep{fastturn2026,jalturn2026} or fold VAD, turn detection, and ASR into one audio-LLM front-end \citep{uaf2026}, architecturally closest to us but with no released weights or recipe. Deployed open detectors run beside the recognizer: Smart Turn (acoustic, open recipe, no transcript; \citealp{smartturn2025}) and the LiveKit and TEN end-of-utterance classifiers over a separate STT's output \citep{livekit2025turndetector,ten2025turndetection,livekit2026eou}. Open-weights recognizers with in-model turn signals, Kyutai STT's semantic-VAD head \citep{kyutai2025stt} and NVIDIA's Parakeet-EOU token \citep{nvidia2026parakeeteou}, publish weights but not the labeling procedure or data construction, and neither measures context biasing. Commercial systems have moved the decision into the recognizer itself: Deepgram's Flux emits end-of-turn events from the forward pass \citep{deepgram2026flux}, OpenAI's Realtime API waits longer after a trailing ``ummm'' than after a completed sentence \citep{openai2025semanticvad}, and AssemblyAI ships a universal streaming endpoint \citep{assemblyai2025universalstreaming}. None of these publishes training details. Our contribution is the missing artifact, the recipe, plus evidence that trade-offs this literature reports as intrinsic can be manufactured by non-causal supervision.

\paragraph{Turn-taking in dialogue, and other streaming decisions.} Turn-taking prediction has a long tradition \citep{skantze2021review}, from text-based end-of-turn prediction \citep{ekstedt2020turngpt} to voice-activity projection trained on future windows \citep{ekstedt2022vap}, with human timing evidence from \citet{stivers2009universals,levinson2015timing}. Projection objectives predict distributions over the future and are causally honest; our diagnosis concerns future-dependent quantities presented as deterministic labels. The same shape appears in simultaneous translation trained with reference-aligned read/write decisions \citep{ma2019stacl} and in tool-call timing for agents cloned from completed trajectories, a streaming sibling of causal confusion in imitation learning \citep{dehaan2019causal}; metric choices can similarly manufacture ``emergence'' \citep{schaeffer2023mirage}.

\paragraph{Streaming and unified ASR.} Streaming recognition spans RNN-T \citep{graves2012sequence}, large weakly supervised models \citep{radford2023whisper}, decoder-only streaming with latency losses \citep{wan2026streaming}, and unified streaming/offline speech-LLMs \citep{qwen3asr2026}; full-duplex speech-to-speech models \citep{defossez2024moshi} dissolve the turn abstraction entirely. We inherit the unified-ASR substrate and add the in-transcript turn event plus the training recipe; vLLM's paged KV \citep{kwon2023vllm} carries our serving path, with attention sinks \citep{xiao2024streamingllm} and bounded-state serving \citep{li2026metronome} as the in-engine efficiency variant of the pinned context prefix.

\paragraph{Context biasing.} Contextual ASR ranges from attention-based deep context \citep{pundak2018deep} to retrieval-scale biasing \citep{brasr2025}, RL-tuned hotwords \citep{hotwordgrpo2025}, and cue-based prompting \citep{commoncues2026}; \citet{contextualearnings2026} concurrently standardizes entity-level evaluation on our corpus; we contribute plain prompt-slot measurements on the open class.

\section{Conclusion}
\label{sec:conclusion}

A voice agent has to know when a turn is over, and the signal that decides it lives in the words, not the silence, so the decision belongs inside the recognizer. This paper gives the first open recipe for building one: one causal labeling rule, two pair constructions, and a deployment-matched benchmark turn a small open model into a single checkpoint that transcribes, detects end-of-turn semantically, handles dictation, and grounds transcription in context, at an operating point no silence timeout reaches. The obstacle was supervision, not architecture: offline corpora encode the future, and the resulting clairvoyant labels manufacture oscillation and phantom trade-offs, diagnosed by intervention and cured, twice, by counterfactual construction. The failure class, and its checklist diagnostic, should transfer to any streaming decision trained from offline logs.

\section*{Acknowledgements}

We thank Mingqi Yang from Minimax AI for early discussions.
This paper was produced using Pine Copilot's voice-directed \emph{whisper coding} workflow~\citep{pineai2026whispercoding}, in which the authors specify, discuss, and review the work by voice while a coding agent (Claude Code with Claude Opus 4.8) carries out the planning, coding, experiments, and paper writing.
We thank BSQL Networking for hosting the NVIDIA RTX PRO 6000 GPU.

\bibliographystyle{plainnat}
\bibliography{refs}

\clearpage

\appendix

\section{Label-specification details}
\label{app:labels}
\begin{figure}[h]
  \centering
  \includegraphics[width=\linewidth]{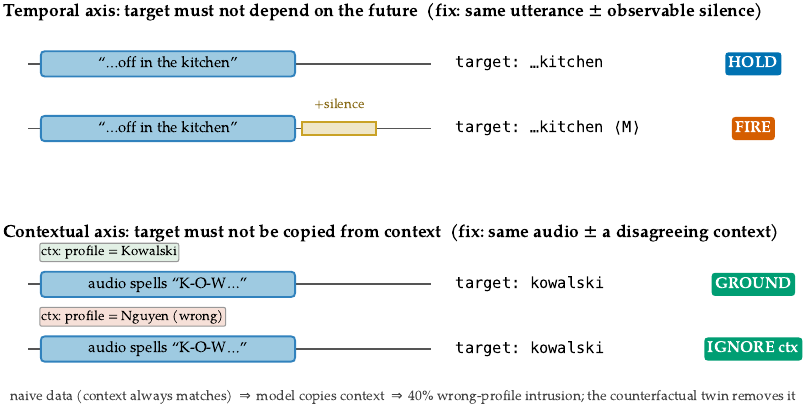}
  \caption{\textbf{One failure class, two instances.} Temporal axis: labels that peek at the future manufacture a phantom frontier; the fix is the minimal pair, the same utterance with and without an observable silence tail. Contextual axis: a context prefix that always matches the audio becomes a copyable shortcut; the fix is the counterfactual twin, the same audio with a matching and a disagreeing context, target following the audio.}
  \label{fig:twoaxes}
\end{figure}

\begin{figure}[h]
  \centering
  \includegraphics[width=\linewidth]{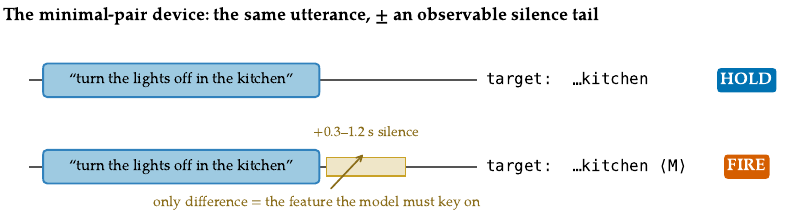}
  \caption{\textbf{The minimal-pair construction.} Each complete utterance appears twice: without a silence tail (target: transcript only) and with a 0.3--1.2\,s tail (target: transcript + marker). The only difference between the opposite-labeled examples is the observable the model must key on.}
  \label{fig:minimalpair}
\end{figure}

The completeness classifier for pair examples (Table~\ref{tab:schemas}, \#4--5) works as follows: utterance A is \emph{incomplete} iff its transcript ends in a filler or connective (``and, but, so, or, uh, um, the, a, to, of, in, with, i mean, you know, \ldots''), ends mid-word (AMI partial-word annotation), or has fewer than 3 words outside a closed acknowledgment list. Borderline cases land in the resume-after-fire cost metric (\S\ref{sec:bench}); they never create fire/hold contradictions on identical observables. Abandonment (incomplete speech, speaker never resumes) is an inference-policy timeout, not a label. Pair examples use both same- and different-speaker AMI pairs: the fire decision keys on completeness + silence, never on speaker identity, which single-channel deployment cannot observe. Figure~\ref{fig:minimalpair} illustrates the minimal-pair construction of \S\ref{sec:recipe}, and Figure~\ref{fig:twoaxes} places it beside its contextual analog, the counterfactual twin of \S\ref{sec:ctxrecipe}.

\section{Benchmark specification and diagnostics}
\label{app:eval}
Full protocol: 0.5\,s chunks; committed-prefix incremental decoding, matching the base model's official streaming semantics (text once emitted stays fixed and is re-fed as the prompt, with the last 5 tokens rolled back so the model can revise them); fires timestamped at the chunk boundary where \fire{} first appears; recall window $[-0.25,+1.5]$\,s, clipped at the next utterance's onset. Set sizes: development 25 stretches (96 turn-final boundaries), post-development confirmation 50 stretches (184 boundaries), held-out 100 stretches (384 boundaries). The first version of the boundary classifier promoted a boundary to turn-final when another meeting speaker interleaved before the target speaker resumed. On a single-channel stream that speech is silence; demanding a fire there is exactly the clairvoyance the benchmark exists to remove. The corrected classification moved the mixed-pools model's recall $0.917{\to}0.906$ and the opposed-pools model's $0.935{\to}0.938$; no conclusion changed, but the episode is a working example of the \S\ref{sec:discussion} checklist applied to our own evaluation.

\begin{figure}[h]
  \centering
  \includegraphics[width=0.55\linewidth]{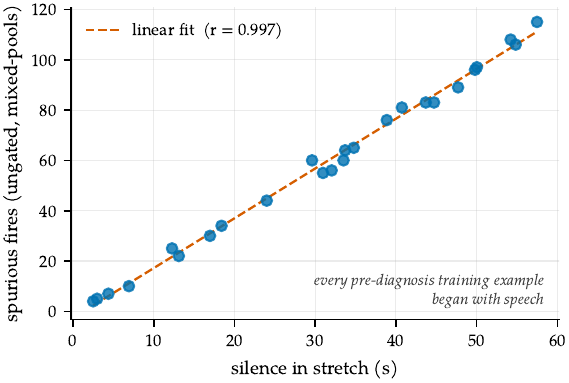}
  \caption{\textbf{Spurious markers on silence.} Spurious fires of the ungated mixed-pools model track the silence content of each stretch almost perfectly; ungated recall is therefore uninterpretable, and all reported configurations run behind the energy gate.}
  \label{fig:silence}
\end{figure}

\paragraph{Spurious markers on silence, quantified.} Models trained only on speech-initial examples hallucinate on silence: spurious fires of the ungated mixed-pools model track each stretch's silence content almost perfectly ($r{=}0.997$; Figure~\ref{fig:silence}), at ${\sim}1.9$ fires per second of silence. This is why \S\ref{sec:bench} runs every comparison behind the energy gate.

\begin{figure}[h]
  \centering
  \includegraphics[width=0.55\linewidth]{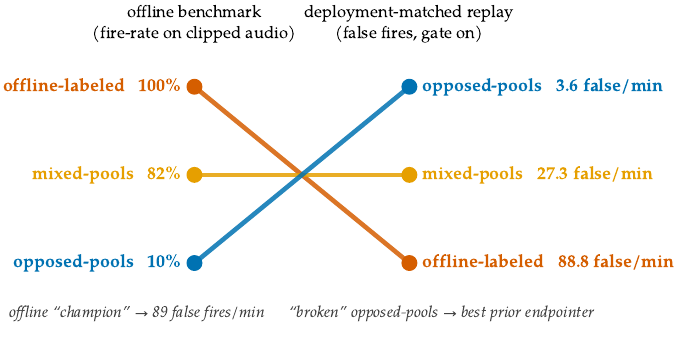}
  \caption{\textbf{The ranking inversion.} Model ranking on the legacy offline benchmark (left) versus the deployment-matched replay benchmark (right). The offline champion is the worst live model.}
  \label{fig:inversion}
\end{figure}

\paragraph{The ranking inversion.} Figure~\ref{fig:inversion} shows the historical model ranking on the legacy offline benchmark versus the deployment-matched replay benchmark (\S\ref{sec:intervention}): the offline champion fires 89 times per speech-minute once audio keeps flowing, while the opposed-pools model, written off at 10\% offline recall, is the best prior endpointer.

\section{The turn-aware system landscape}
\label{app:landscape}

\begin{table}[h]
  \centering\footnotesize
  \caption{The landscape (verified July 2026; discussion in \S\ref{sec:related}). ``Turn detectors'' are separate models beside a recognizer; ``cascade'' is VAD + silence timeout + separate ASR, the deployed default and our measured baseline.}
  \label{tab:matrix}
  \setlength{\tabcolsep}{2.4pt}
  \begin{tabular}{@{}lcccccc@{}}
    \toprule
    & This work & Flux / \texttt{sem.vad} & Kyutai STT & Parakeet-EOU & Turn det. & Cascade \\
    \midrule
    Streaming ASR & \checkmark & \checkmark & \checkmark & \checkmark & -- & \checkmark \\
    In-model end-of-turn & semantic & semantic & sem.\ head & token & sep.\ model & timeout \\
    Open weights & -- & -- & \checkmark & \checkmark & mostly & \checkmark \\
    Turn-training recipe & \checkmark & -- & -- & -- & Smart Turn & n/a \\
    Context biasing, measured & \checkmark\,($+28.9$\,pp) & -- & -- & -- & -- & ASR-dep. \\
    \bottomrule
  \end{tabular}
\end{table}

\section{Additional results}
\label{app:tables}

\paragraph{Gated configurations and the confirm-horizon sweep.} Table~\ref{tab:appconfirm}. The confirm horizon behaves as a clean latency-vs-false-fire dial: $h{=}1$ strictly improves the mixed-pools model (recall \emph{rises} because deferring the fire lets the segment gather the boundary inside tolerance) and zeroes the opposed-pools and causal models' false fires; $h{=}2$ over-suppresses (mixed-pools recall 0.812, latency past budget). The causal model is the only checkpoint whose $h{=}0$ row is already deployable. Ungated configurations are omitted as uninterpretable (Figure~\ref{fig:silence}): with ${\sim}1.9$ spurious fires per second on silence, the pre-diagnosis models score 0.95--0.99 ``recall'' by chance.

\begin{table}[h]
  \centering\small
  \caption{All gated configurations on the 25-stretch development benchmark. Resume = resume-after-fire rate over continuation boundaries (a cost metric, not an error rate). WER is a fraction, unbounded above through insertions.}
  \label{tab:appconfirm}
  \begin{tabular}{lcccccc}
    \toprule
    Configuration & Recall & P50 & P95 & False/min & Resume & WER$_\text{mean}$ \\
    \midrule
    offline-labeled + gate & 0.896 & 0.05\,s & 0.25\,s & 88.8 & 0.89 & 0.85 \\
    mixed-pools + gate & 0.906 & 0.16\,s & 0.89\,s & 27.3 & 0.96 & 0.36 \\
    mixed-pools + gate + confirm $h{=}1$ & 0.927 & 0.68\,s & 1.27\,s & 3.2 & 0.85 & 0.36 \\
    mixed-pools + gate + confirm $h{=}2$ & 0.812 & 1.15\,s & 1.42\,s & 1.1 & 0.56 & 0.36 \\
    opposed-pools + gate & 0.938 & 0.26\,s & 0.56\,s & 3.6 & 0.93 & 4.96 \\
    opposed-pools + gate + confirm $h{=}1$ & 0.938 & 0.77\,s & 1.06\,s & 0.0 & 0.85 & 4.96 \\
    causal + gate & 0.969 & 0.39\,s & 0.70\,s & 0.3 & 1.00 & 1.43 \\
    causal + gate + confirm $h{=}1$ & 0.969 & 0.89\,s & 1.20\,s & 0.0 & 0.93 & 1.43 \\
    \bottomrule
  \end{tabular}
\end{table}

\paragraph{The full timeout family.} Table~\ref{tab:apptimeout}, both detectors on the development (25 stretches) and fresh confirmation (50 stretches) sets. The family's shape is stable across sets: recall collapses between $X{=}1.0$ and $1.5$\,s (the fire lands outside the $+1.5$\,s tolerance), and no setting reaches the causal model's (recall, latency, false-fire) point; matching its recall costs a timeout $3\times$ the latency and ${\sim}5\times$ the false fires. Silero \citep{silero2021vad} ran per-chunk with state resets, a slight handicap; RMS shares the LM configurations' exact detector and is the apples-to-apples policy comparison; on this close-talking microphone channel the Silero cascade is dominated by plain RMS.

\begin{table}[h]
  \centering\small
  \caption{Silence-timeout cascade, all settings, on the development (dev) and fresh confirmation (conf) stretch sets.}
  \label{tab:apptimeout}
  \begin{tabular}{llcccccc}
    \toprule
    Detector & Set & $X$ (s) & Recall & P50 & P95 & False/min & Resume \\
    \midrule
    RMS & dev & 0.5 & 0.958 & 0.65\,s & 0.94\,s & 6.3 & 1.00 \\
    RMS & dev & 1.0 & 0.969 & 1.15\,s & 1.44\,s & 1.4 & 0.81 \\
    RMS & dev & 1.5 & 0.208 & 1.38\,s & 1.49\,s & 0.8 & 0.07 \\
    RMS & dev & 2.0 & 0.031 & 0.55\,s & 0.55\,s & 0.2 & 0.00 \\
    RMS & conf & 0.5 & 0.978 & 0.70\,s & 0.95\,s & 4.6 & 0.99 \\
    RMS & conf & 1.0 & 0.989 & 1.20\,s & 1.45\,s & 0.6 & 0.73 \\
    RMS & conf & 1.5 & 0.185 & 1.44\,s & 1.50\,s & 0.4 & 0.07 \\
    RMS & conf & 2.0 & 0.016 & 1.14\,s & 1.14\,s & 0.2 & 0.00 \\
    Silero & dev & 0.5 & 0.844 & 0.54\,s & 0.88\,s & 13.5 & 0.96 \\
    Silero & dev & 1.0 & 0.854 & 1.04\,s & 1.38\,s & 5.5 & 0.85 \\
    Silero & dev & 1.5 & 0.417 & 1.24\,s & 1.49\,s & 2.5 & 0.41 \\
    Silero & dev & 2.0 & 0.125 & 1.32\,s & 1.42\,s & 1.0 & 0.00 \\
    Silero & conf & 0.5 & 0.837 & 0.51\,s & 0.83\,s & 12.3 & 0.88 \\
    Silero & conf & 1.0 & 0.864 & 1.00\,s & 1.33\,s & 4.4 & 0.73 \\
    Silero & conf & 1.5 & 0.451 & 1.29\,s & 1.48\,s & 2.0 & 0.27 \\
    Silero & conf & 2.0 & 0.130 & 1.34\,s & 1.44\,s & 1.3 & 0.07 \\
    \bottomrule
  \end{tabular}
\end{table}

\paragraph{Fresh-set confirmation.} The causal model with the gate on the confirmation set: recall 0.924, P50 0.42\,s, P95 0.70\,s, 0.21 false fires per speech-minute, resume 0.91, WER median 0.30 / mean 4.63 (three long-monologue stretches enter the no-flush repetition loop; the force-flush of Appendix~\ref{app:serving} is the documented fix). The 4.5\,pp recall drop from the development set (93/96 vs.\ 170/184 turn-final boundaries) is consistent with sampling variation: the two-proportion 95\% interval on the difference, $[-0.7,+9.7]$\,pp, includes zero.

\paragraph{Probe confidence intervals.} On the enlarged probes (250 dictation sequences; 240 spelled items over 120 identities), the unified model's 95\% intervals are: premature fires per sequence $0.16$ [$0.12,0.22$], final recall $0.88$ [$0.84,0.92$], email-with-context $0.93$ [$0.87,0.97$], intrusion $0.008$ [$0.002,0.030$]. On the 100-stretch replay set: boundary recall $0.982$ [$0.965,0.995$] at $1.30$ [$0.91,1.71$] false fires per speech-minute.

\paragraph{Dictation scoring-window sensitivity.} The shipped scoring window counts any fire within $+0.25$\,s of the last speech end as premature and denies it final-recall credit, which is strict for a low-latency endpointer. Re-scoring every fire by its offset $\delta$ from the true end of speech (genuine mid-sequence premature: $\delta<-0.30$\,s; on-time: $-0.30\le\delta\le+1.75$\,s; late: $\delta>+1.75$\,s) gives, on the same 250 sequences: 0.07 genuine mid-sequence fires per number (18/250 sequences), 0.968 on-time final recall, zero late fires, and end-fire latency P50 0.44\,s (fires are quantized up to the 0.5\,s chunk grid). The residual under the shipped window is therefore eagerness within one chunk of the true end, not lateness; Table~\ref{tab:dictation} reports the shipped window for comparability across rows.

\paragraph{Unified-model endpointing across sets.} Table~\ref{tab:released} collects the unified rank-32 checkpoint's replay numbers, quoted piecewise in \S\ref{sec:released}, next to the pure-endpointing causal checkpoint of \S\ref{sec:endpointresults}. The breadth-versus-precision cost (0.97 versus 0.3 false fires per speech-minute) is a development-set comparison at matched latency; a one-chunk confirm horizon zeroes the unified model's false fires at 0.958 recall (P50 0.89\,s) for deployments that prefer conservatism over latency, and the added schemas leave endpointing recall essentially intact at rank 32 (0.95 versus 0.97 on the development set).

\begin{table}[h]
  \centering\small
  \caption{The unified checkpoint (rank-32) on the replay benchmark, both stretch sets, alongside the pure-endpointing causal checkpoint of \S\ref{sec:endpointresults}. dev = 25-stretch development set; held-out 100 = the 100-stretch set; fresh 50 = the post-development confirmation set.}
  \label{tab:released}
  \begin{tabular}{llcccc}
    \toprule
    Model & Set & Recall & P50 & P95 & False/min \\
    \midrule
    unified r32 + gate & dev & 0.948 & 0.39\,s & 0.65\,s & 0.97 \\
    unified r32 + gate + confirm $h{=}1$ & dev & 0.958 & 0.89\,s & 1.15\,s & 0.0 \\
    unified r32 + gate & held-out 100 & 0.982 & 0.38\,s & 0.64\,s & 1.30 \\
    \midrule
    pure endpoint (causal, r16) + gate & dev & 0.969 & 0.39\,s & 0.70\,s & 0.3 \\
    pure endpoint (causal, r16) + gate & fresh 50 & 0.924 & 0.42\,s & 0.70\,s & 0.2 \\
    \bottomrule
  \end{tabular}
\end{table}

\paragraph{Seed robustness of the unified recipe.} Table~\ref{tab:seeds}. Retraining the unified recipe (rank-32 adapter, unified ${\sim}20$k pool with 20\% plain-ASR replay and natural-speech biasing, weight-decay and cosine fixes) under three seeds and selecting each best checkpoint on the probe axes gives stable premature-fire, context, and intrusion numbers; the unified capability is a property of the recipe, not a lucky checkpoint.

\begin{table}[h]
  \centering\small
  \caption{Unified recipe under three training seeds (best checkpoint each, smaller common probe set; the unified checkpoint's enlarged-probe figures, email $0.93$ and intrusion $0.8\%$, are in Table~\ref{tab:dictation}). Seed 0 is the unified checkpoint.}
  \label{tab:seeds}
  \begin{tabular}{lccc}
    \toprule
    Seed & Premature/seq\,$\downarrow$ & Email $+$ctx\,$\uparrow$ & Intrusion\,$\downarrow$ \\
    \midrule
    0 (unified) & 0.16 & 0.88 & 0.000 \\
    1 & 0.16 & 0.98 & 0.000 \\
    2 & 0.16 & 0.85 & 0.000 \\
    \midrule
    mean\,$\pm$\,sd & $0.16\pm0.00$ & $0.90\pm0.07$ & $0.000\pm0.000$ \\
    \bottomrule
  \end{tabular}
\end{table}

\paragraph{Schema ablation.} Table~\ref{tab:ablation}. We retrain the endpointing recipe with one construction removed at a time, keeping data volume fixed by redistributing the dropped quota across the remaining schemas; each configuration is selected on the holdout composite (mean accuracy across the eight schema classes on held-out data) and scored on the 25-stretch replay set. Each construction has a distinct predicted failure; two of three reproduce it. Removing the \textbf{minimal pair} (schema 2) makes silence optional again: gated false fires rise $1.08{\to}5.28$ per speech-minute and the confirm horizon cancels 34 candidates versus 10, so both the pathology and the reliance on the confirm horizon return. Removing the \textbf{silence schemas} (7--8) costs competence on silence: without the gate, the model emits 1458 markers on silence versus 8; the energy gate masks this in the gated numbers. Removing the \textbf{pause pair}'s incomplete half (schema 5) does not raise false fires; the model is mildly more conservative (recall $0.948{\to}0.906$, false fires $1.08{\to}0.43$), so pause discrimination is largely covered by schema 4's complete-A gaps.

\begin{table}[h]
  \centering\small
  \caption{Schema ablation on the endpointing pool (data volume held fixed; 25-stretch replay, best checkpoint each). Each configuration is an independent retrain selected on the holdout composite, so the baseline here (0.948/1.08) is a different checkpoint from the causal model of Table~\ref{tab:main} (0.969/0.3). ``Silence-fires'' is the ungated marker count on silence; ``$h{=}1$ cancels'' counts the fire candidates the confirm horizon must cancel; a high count means the model still relies on it.}
  \label{tab:ablation}
  \begin{tabular}{lccccc}
    \toprule
    Configuration & Recall\,$\uparrow$ & False/min\,$\downarrow$ & Silence-fires\,$\downarrow$ & $h{=}1$ cancels & Predicted failure? \\
    \midrule
    baseline (full pool) & 0.948 & 1.08 & 8 & 10 & -- \\
    $-$minimal pair (\#2) & 0.906 & 5.28 & 5 & 34 & yes (silence optional) \\
    $-$silence (\#7--8) & 0.969 & 1.94 & 1458 & 12 & yes (silence spray) \\
    $-$pause pair (\#5) & 0.906 & 0.43 & 8 & 4 & no (redundant) \\
    \bottomrule
  \end{tabular}
\end{table}

\paragraph{Offline WER regression.} Table~\ref{tab:wer}. The fine-tune costs $+1.1$/$+2.7$\,pp on the full official LibriSpeech splits (single-shot decode on vLLM, jiwer/whisper normalization). ``Fires'' is the fraction of clips where the model emitted the endpoint marker; offline clips end at or near end-of-speech, and the base model never fires because its marker rows are untrained. Banning the markers at decode time zeroes the fires without moving WER, which rules out decode truncation as the cause; Appendix~\ref{app:recipe} attributes the regression to narrow-schema drift and quantifies the plain-ASR-replay mitigation. The unified model, which folds a 20\% plain-ASR-replay fraction into a rank-32 adapter (\S\ref{sec:ctxrecipe}), recovers most of the gap, scoring 3.5\%/6.9\%.

\begin{table}[h]
  \centering\small
  \caption{Offline WER (LibriSpeech, full official splits).}
  \label{tab:wer}
  \begin{tabular}{lcccc}
    \toprule
    & \multicolumn{2}{c}{WER\,$\downarrow$} & \multicolumn{2}{c}{Fires} \\
    \cmidrule(lr){2-3}\cmidrule(lr){4-5}
    Model & clean & other & clean & other \\
    \midrule
    Qwen3-ASR-0.6B (base) & 2.79\% & 5.14\% & 0\% & 0\% \\
    + endpoint LoRA (merged, causal, endpoint-only) & 3.93\% & 7.87\% & 4.7\% & 5.9\% \\
    + markers banned at decode & 3.94\% & 7.91\% & 0\% & 0\% \\
    \textbf{unified (r32, $+$replay)} & \textbf{3.49\%} & \textbf{6.90\%} & 3.2\% & 4.7\% \\
    \bottomrule
  \end{tabular}
\end{table}

\section{Diagnosis details}
\label{app:markerprob}

\begin{figure}[h]
  \centering
  \includegraphics[width=\linewidth]{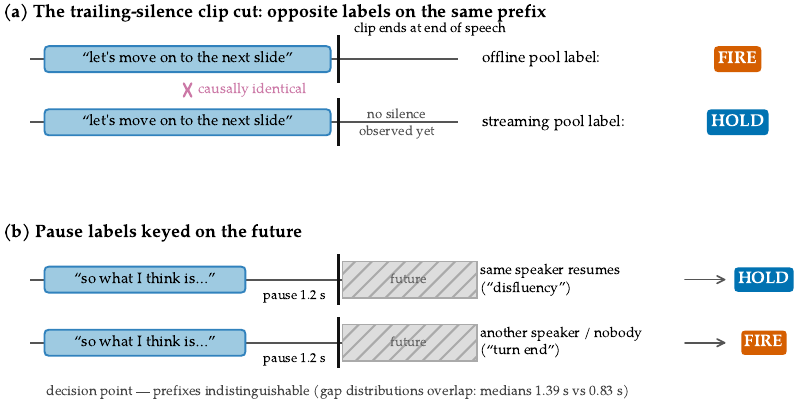}
  \caption{\textbf{Two ways offline-derived supervision leaks the future.} (a)~Offline clips inherit forced-alignment cuts at end of speech, so the offline pool demands a fire on exactly the prefix the streaming pool labels \emph{hold}. (b)~Disfluency-vs-boundary pause labels are keyed on who speaks next; at the decision point the two classes' silence-gap distributions overlap almost completely.}
  \label{fig:contradictions}
\end{figure}

\paragraph{The two leaks, illustrated.} Figure~\ref{fig:contradictions} shows the two mechanisms of \S\ref{sec:intervention} side by side: the forced-alignment clip cut that puts opposite labels on the same prefix, and the next-speaker pause labels whose classes are indistinguishable at the decision point (gap medians 1.39\,s for disfluent pauses vs.\ 0.83\,s for turn boundaries).

\begin{figure}[h]
  \centering
  \includegraphics[width=0.6\linewidth]{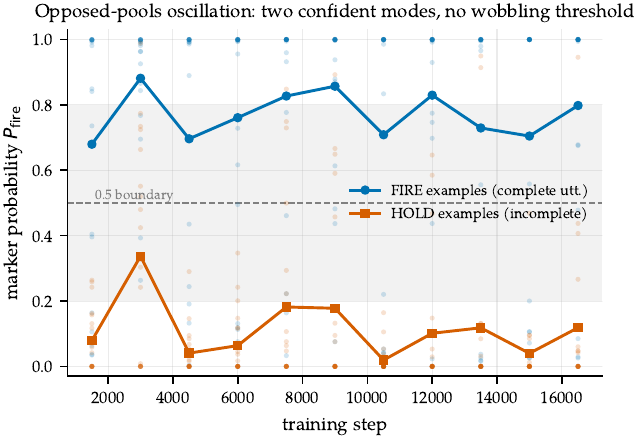}
  \caption{\textbf{Two modes in the weights, not a wobbling threshold.} Mean marker-commit probability $P_{\text{fire}}$ for fire-labeled (complete utterance) and hold-labeled (incomplete) holdout examples across the opposed-pools run; small dots are per-example values. The two classes swing in lockstep as the weights cycle between the fire mode and the no-fire mode; no example hovers persistently at the 0.5 decision boundary.}
  \label{fig:markerprob}
\end{figure}

\paragraph{The probability-level oscillation probe.} Figure~\ref{fig:markerprob} is the measurement behind the mode-cycling claim of \S\ref{sec:analysis}. The trained marker pathway is a rigid two-token sequence (\eager{} then \fire{}), deterministic once entered, so we define $P_{\text{fire}}$ as the probability that the model's continuation enters the marker path rather than terminating. It is evaluated at the end-of-audio decision position with the transcript teacher-forced, marginalizing over the first continuation token at ${\geq}0.98$ probability coverage and greedy-rolling each branch.

Two observations follow. On the balanced holdout of the opposed-pools run (the exact examples behind Figure~\ref{fig:oscillation}, left), mean $P_{\text{fire}}$ tracks the recorded accuracy ($r{=}{+}0.98$ on the contested fire schema, $r{=}{-}0.95$ against hold-class accuracy) and swings with amplitude ${\sim}0.4$ across snapshots. And the per-example distribution rules out a wobbling decision threshold: not one of the 40 held-out examples sits in the ambiguous $[0.2,0.8]$ band across all eleven snapshots (only 13--15\% sit there at any single snapshot), while 7 of 40 fully reverse between confident fire (${>}0.8$) and confident hold (${<}0.2$) as training proceeds. The trailing-silence schema, on which both pools agree, stays pinned at $P_{\text{fire}}{\approx}1.0$ throughout: the cycling is confined to the schemas the pools contest.

\section{The synthetic clairvoyant-fraction experiment}
\label{app:toy}
Setup for the \S\ref{sec:toy} experiment: a discrete stream of token phrases separated by silence runs, with no acoustics and no pretraining. A phrase ends complete or incomplete, always readable from the prefix, while each gap's full length (speaker resumes vs.\ turn ends, 50/50) is knowable only after the decision point. At a fixed depth into every gap, a two-layer causal transformer (${\sim}30$k parameters) must emit fire or hold. The \emph{causal} rule fires iff the phrase was complete; the \emph{clairvoyant} rule fires iff the gap turns out long, a genuine future dependence and the direct analog of clips cut at boundaries and transcripts recording who spoke next. Completeness and gap length are independent, so the two rules are maximally opposed while each remains self-consistent, reproducing the structure of \S\ref{sec:principle}: pools that individually make sense yet demand opposite behavior on the same input. A fraction $f$ of training sequences is clairvoyantly labeled; the holdout is always causal; three seeds per $f$; everything runs on CPU in minutes.

\begin{figure}[h]
  \centering
  \includegraphics[width=\linewidth]{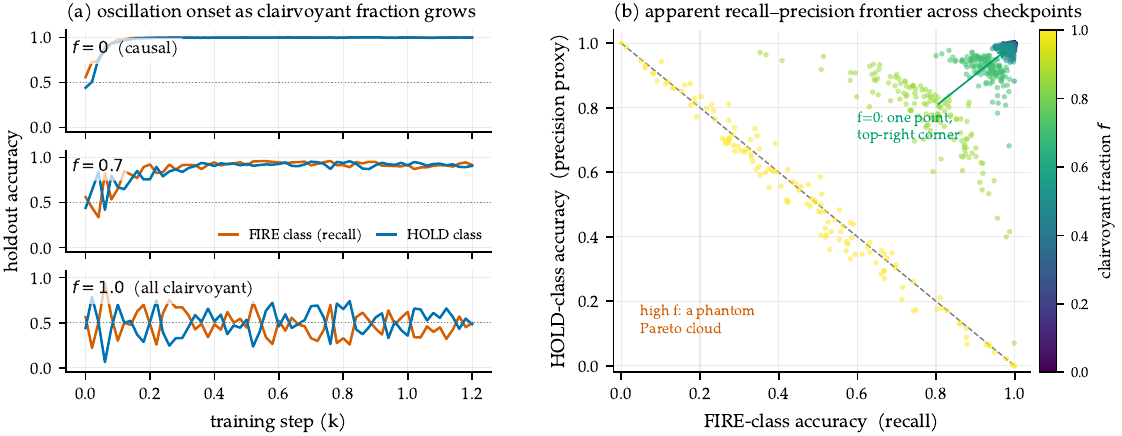}
  \caption{\textbf{The mechanism, isolated.} A ${\sim}30$k-parameter transformer on a synthetic stream with a tunable clairvoyant-label fraction $f$. (a)~Holdout trajectories: monotone convergence at $f{=}0$, visible noise at $f{=}0.7$, oscillation between two modes at $f{=}1.0$. (b)~Every checkpoint's (fire recall, hold precision), colored by $f$: the causal run collapses to one top-right point; high $f$ spreads checkpoints into an anti-diagonal cloud, a phantom Pareto frontier traced by one oscillating model.}
  \label{fig:toy}
\end{figure}

The dose-response in detail (Figure~\ref{fig:toy}): at $f{=}0$, both classes reach 1.00, zero mode flips, worst post-warmup holdout accuracy 0.99. The fire/hold accuracy anti-correlation grows monotonically ($-0.3$ at $f{=}0.1$, $-0.5$ at $f{=}0.5$, $-1.0$ at $f{=}1.0$), while outright mode flipping appears only at high fraction: 8 flips per run at $f{=}0.85$ and 16 at $f{=}1.0$, zero at $f{\leq}0.7$.

\section{Training-recipe notes}
\label{app:recipe}
All models share one adapter configuration: LoRA on the attention q/k/v/o projections plus the two marker embedding rows, $\alpha{=}2r$, batch 8 ($r{=}16$ for the endpointing experiments, $r{=}32$ for the unified checkpoint, which is selected on the probe axes of dictation, context, and intrusion rather than the holdout composite alone). The unified model (\S\ref{sec:ctxrecipe}) adds two AdamW hygiene fixes and one negative result. All three were isolated on identical data (the unified pool), so the comparisons below vary only the optimizer setting.

\paragraph{A weight-decay bug worth fixing, but not the WER culprit.} We reserve two vocabulary rows as markers and train them by unfreezing the (tied) token-embedding and output-head matrix and masking gradients to those two rows. But AdamW's \emph{decoupled} weight decay applies $p \leftarrow p - \eta\lambda p$ to \emph{every} row regardless of gradient, so at $\eta{=}2\!\times\!10^{-4},\lambda{=}0.01$ the entire (tied) 151k-row matrix shrinks ${\sim}2.4\%$ over 12k steps, silently degrading the base model's vocabulary and output head. This is a general hazard for any recipe that adds a few trainable vocabulary rows to a frozen embedding, and we fix it with a $\lambda{=}0$ group. The negative half: in a controlled comparison on identical data, the fix (plus cosine schedule) did \emph{not} lower offline WER; the fixed unified checkpoint scores 3.9\%/7.3\% (clean/other) versus 3.6\%/7.0\% without it, within checkpoint-selection noise and in the wrong direction. The offline-WER regression is therefore \emph{not} caused by the weight-decay bug; it is narrow-schema drift from fine-tuning on ${\sim}12$k examples with no plain-ASR replay data. We keep the fix for its base-model-preservation rationale, not for a WER win we cannot demonstrate.

\paragraph{Plain-ASR replay: the fix that does move WER.} The unified model mixes ${\sim}4$k plain transcription examples (audio $\to$ transcript, no marker, no context) into the ${\sim}20$k-example pool, a 20\% fraction. They teach the model that a complete utterance with no context is simply to be transcribed, recovering offline WER from 3.9\%/7.3\% (recipe fixes, no replay) to 3.5\%/6.9\% and cutting spurious offline marker fires from ${\sim}10$--$20\%$ to 3.2\%. The replay fraction trades against endpointing sharpness at fixed adapter capacity; raising the LoRA rank from 16 to 32 buys back that sharpness, which is why the unified checkpoint is rank-32 (\S\ref{sec:released}).

\paragraph{Cosine decay.} A cosine schedule from $2\!\times\!10^{-4}$ to $10^{-5}$ after a 100-step warmup gives the most stable, highest holdout-composite trajectory (plateau $0.992$ from step 6k) and a slightly less eager final checkpoint than the constant schedule.

\paragraph{Muon underperforms AdamW here.} We tried Muon \citep{jordan2024muon} on the 2-D LoRA factors (Newton-Schulz-orthogonalized momentum, AdamW retained for the marker rows) at a $100\times$ larger base LR ($2\!\times\!10^{-2}$, since the orthogonalized update is ${\sim}$unit-norm). On identical data Muon's raw loss descends \emph{faster} early but its task metric plateaus at holdout composite $0.846$ (by step 3--4.5k, conversational schema collapsing) versus AdamW's $0.985$. The cause is structural: rank-16 LoRA factors ($1024\times16$) are too rectangular for the 5-step Newton-Schulz iteration to orthogonalize (empirically $\|O^\top O - I\|$ off-diagonals ${\approx}0.5$), and orthogonalizing the two factors separately is not the same operation as orthogonalizing their product. We keep AdamW.

\section{Serving details}
\label{app:serving}
\begin{figure}[h]
  \centering
  \includegraphics[width=\linewidth]{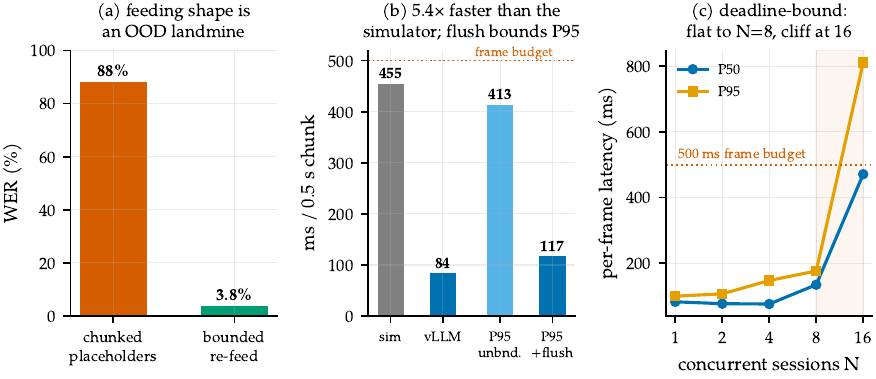}
  \caption{\textbf{Serving on real infrastructure.} (a)~Feeding shape: per-chunk placeholders are out-of-distribution; bounded re-feed is correct. (b)~Per-chunk compute on vLLM vs.\ the transformers simulator; the force-flush bounds the P95 tail. (c)~Concurrent sessions per engine on a 0.15-GPU slice: flat to $N{=}8$, cliff at 16.}
  \label{fig:serving}
\end{figure}

Three findings from running the model on shared-GPU vLLM \citep{kwon2023vllm} rather than a simulator (Figure~\ref{fig:serving}), summarized in \S\ref{sec:released}. First, the feeding shape matters. The natural flat-cost shape, appending each 0.5\,s chunk as a new audio placeholder in a resident request, is out of distribution for a model trained on single segments and destroys it (88\% WER); re-feeding the last bounded window of audio each chunk is the correct primitive (3.8\%). Second, the vLLM path reproduces the endpoint behavior at 84\,ms median per chunk, $5.4\times$ faster than the transformers simulator used during development; latency claims made on $O(T^2)$ re-decode simulators overstate the cost of this approach by that factor.

Third, bounding every resident state matters: a 20\,s force-flush fixes both the WER tail and the compute tail, and per-frame P95 stays flat through 8 concurrent sessions on a 0.15-GPU slice (a lower bound measured under contention; the shape, not the constant, is the result). Re-measured on the unified rank-32 checkpoint, architecturally identical to the endpoint model after merging, serving is unchanged: median per-chunk compute is 21\,ms on an otherwise-idle GPU (the 84\,ms above is under a co-tenant), and per-frame P95 stays flat through 16 concurrent sessions (77\,ms at $N{=}16$, well inside the 500\,ms budget).

\end{document}